\documentclass[slac_one]{revtex4}
\usepackage{graphicx}
\usepackage{fancyhdr}
\begin{document}

\title{Commissioning ATLAS Trigger}

\author{T. Bold on behalf of ATLAS TDAQ}
\affiliation{AGH-University of Science and Technology, Krakow, Poland \& University of California Irvine, Irvine CA, US }

\begin{abstract}
The ATLAS experiment at the Large Hadron Collider (LHC) will face the
challenge of efficiently selecting interesting candidate events in $pp$
collisions at 14\,TeV centre-of-mass energy, whilst rejecting the
enormous number of background events. Therefore it is equipped with a three level
trigger system. The first level is is hardware based and uses coarse granularity
calorimeter information and fast readout muon chambers.
The second and third level triggers, which are software based,
will need to reduce the first level trigger output rate of $\sim$\,75\,$\mathrm{kHz}$ to
$\sim$\,200\,$\mathrm{Hz}$ written out to mass storage.
The progress in commissioning of this system will be reviewed in this paper.
\end{abstract}

\maketitle

\section{The ATLAS trigger system}

\begin{figure}[!ht]
\centering
\includegraphics[width=3.3in]{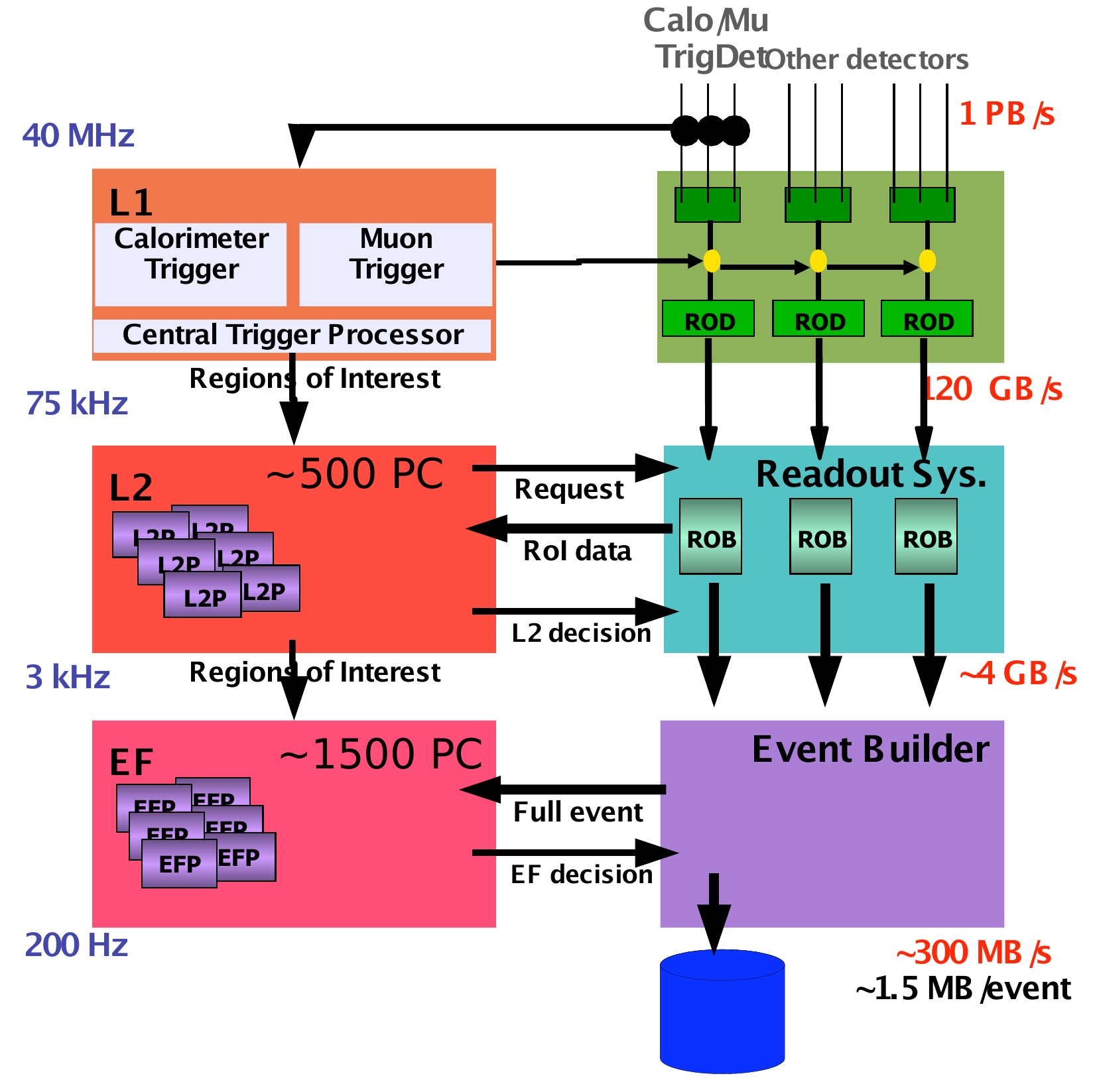}
\caption{Sketch of the ATLAS T/DAQ system. The right side boxes show the data collection infrastructure while the left side show trigger components.
Abbreviations: L1 - first level trigger, L2 - second level trigger, EF - third level trigger (known as Event Filter), ROD/ROB - readout driver/buffer respectively, 
L2P EFP - are L2 processes and EF processes respectively. Multiple boxes are used to express the fact that L2 and EF consist of farms of PCs. }
\label{fig:tdaq_figure}
\end{figure}

The Large Hadron Collider (LHC) at CERN, Geneva, is now starting
operation. It will ultimately provide proton-proton collisions at a
centre-of-mass energy of $14\,\mathrm{TeV}$, a design luminosity of
$10^{34}\,\mathrm{cm}^{-2}\,\mathrm{s}^{-1}$ and a bunch-crossing rate
of $40\,\mathrm{MHz}$. The ATLAS collaboration has built a general
purpose experiment for the LHC which is described in
\cite{ATLAS:performaceTDR1}\cite{ATLAS:performaceTDR2}. The trigger
and data acquisition (T/DAQ) system must work in the challenging
environment of $\sim 10^9$ interactions per second and the large
number ($\sim 10^8$) of readout channels of the ATLAS detector. The
initial data stream of $1 \mathrm{PB/s}$ must be reduced to the $\sim
300\,\mathrm{MB/s}$ which can be sustained to mass storage, while
efficiently retaining a maximum acceptance of physics signatures for offline analysis. To
achieve this, ATLAS has designed a three-level trigger system (see
Fig.~\ref{fig:tdaq_figure}) \cite{Jenni:616089}.

The first level trigger (LVL1) is implemented in custom electronics
(mainly ASICs and FPGAs). Its decision is based on relatively coarse data
from two subsystems, the calorimeters and dedicated muon trigger
stations. Events are selected based on inclusive high-$p_t$ objects
(muons, electromagnetic/tau/hadronic
clusters, jet clusters) plus global event features (missing and scalar transverse energy sums) There are a number of programmable trigger thresholds 
for each of these.
During the LVL1 latency of $2.5\,\mathrm{\mu s}$ the data of all sub-detectors are
kept in pipeline memories. For accepted events, the geometrical location
of the objects, {\it Regions of Interest (RoIs)}, are sent to the second
level trigger (LVL2) and the data are then transferred from the pipeline
memories to the Read-Out Buffers (ROBs). The LVL1 trigger reduces the event rate from
the initial $40\,\mathrm{MHz}$ to about $75\,\mathrm{kHz}$.

The High-Level-Trigger (HLT) is a software-based trigger, running on farms
built from commodity computing and network technology. It is
subdivided into LVL2 and the Event Filter (EF). LVL2 has a nominal
average processing time of $\sim 40\,\mathrm{ms}$ and should reduce the output rate to
around $2\,\mathrm{kHz}$. The EF can take around $4\,\mathrm{s}$ and
should further reduce the rate to $\sim 200\,\mathrm{Hz}$. Both levels
have access to the full granularity of all the detector data and
follow the principle of further refining the signatures identified at
LVL1.  LVL2 must retrieve event fragments from the ROBs via
Ethernet. To reduce the data transfer to a few percent, it uses only
data in RoIs identified by LVL1.  LVL2 algorithms are highly optimized
for speed. If LVL2 accepts an event, all the fragments from the ROBs
are combined and sent to one EF processor for further
consideration. The EF further refines the classification of LVL2,
using the extra time to run more complex algorithms, often based on
the same tool set as offline reconstruction. It also benefits from
more detailed calibration and alignment than used at LVL2. The processing at
the EF is based mainly on the RoIs however the full detector
information can be accessed and this capability is used, for example, 
in triggers involving missing transverse energy.

\section{The trigger menu}
The overall configuration of the trigger is called a menu. It is composed of
building blocks, called trigger chains, which can be considered as 
the units of
selection in that the event is accepted if at least one trigger chain is
passed.  Examples of trigger chains are the identification 25 $GeV$ electrons 
or $6
GeV$ muons etc.  This modular structure greatly simplifies the configuration of the
trigger and allows for great flexibility as specific chains can be added
or removed to the menu easily.  The rate can be also controlled
chain-wise by the use of prescaling - this means that a given chain is only
run for a specified fraction of events chosen randomly, effectively reducing the rate for that chain by the prescale factor.

Such decomposition of the whole trigger selection into chains facilitates
the tuning of the trigger selection to adapt to the beam and detector 
conditions as well as to the 
overall ATLAS experimental program. 

\begin{figure}[!ht]
\centering
\includegraphics[width=2.5in]{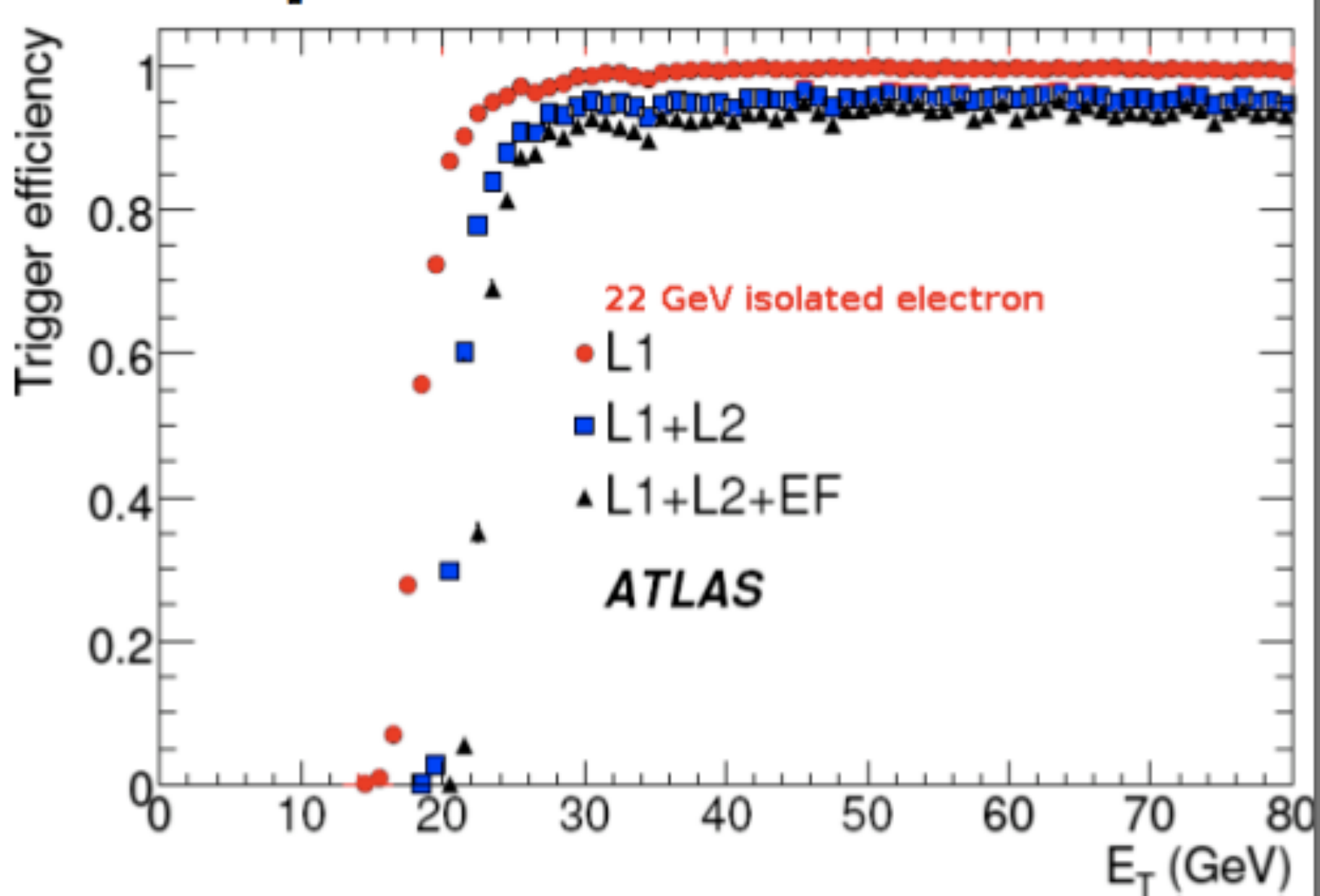}
\caption{The trigger turn-on curve for the 22 GeV electron chain after each of the 3 trigger levels. This plot is based on simulated data i.e. with the use of ``truth'' information. The efficiency determination from data is also studied.}
\label{fig:e22i}
\end{figure}

Work on the menu is divided into working groups based arround the ATLAS 
sub-detectors and the event-features of interest for trigger selection e/$\gamma$, $\tau$, jets, $\mu$, missing-ET, $b$-jet, B-physics \cite{AtlasDetPaper:2008}. These groups perform detailed performance optimizations, 
an example plot showing the efficiencies for single trigger chain are shown in Fig.~\ref{fig:e22i}

This work of the individual working groups is integrated into a set of
trigger menus adapted to  different phases of the experiment. The main 
consideration
for these menus is to provide a full coverage of the physics programme
within the limitations of the maximum rate-to-tape which DAQ system can sustain and the offline limitations for data processing and storage.

\begin{figure}[!ht]
\centering
\includegraphics[width=2.5in, angle=90]{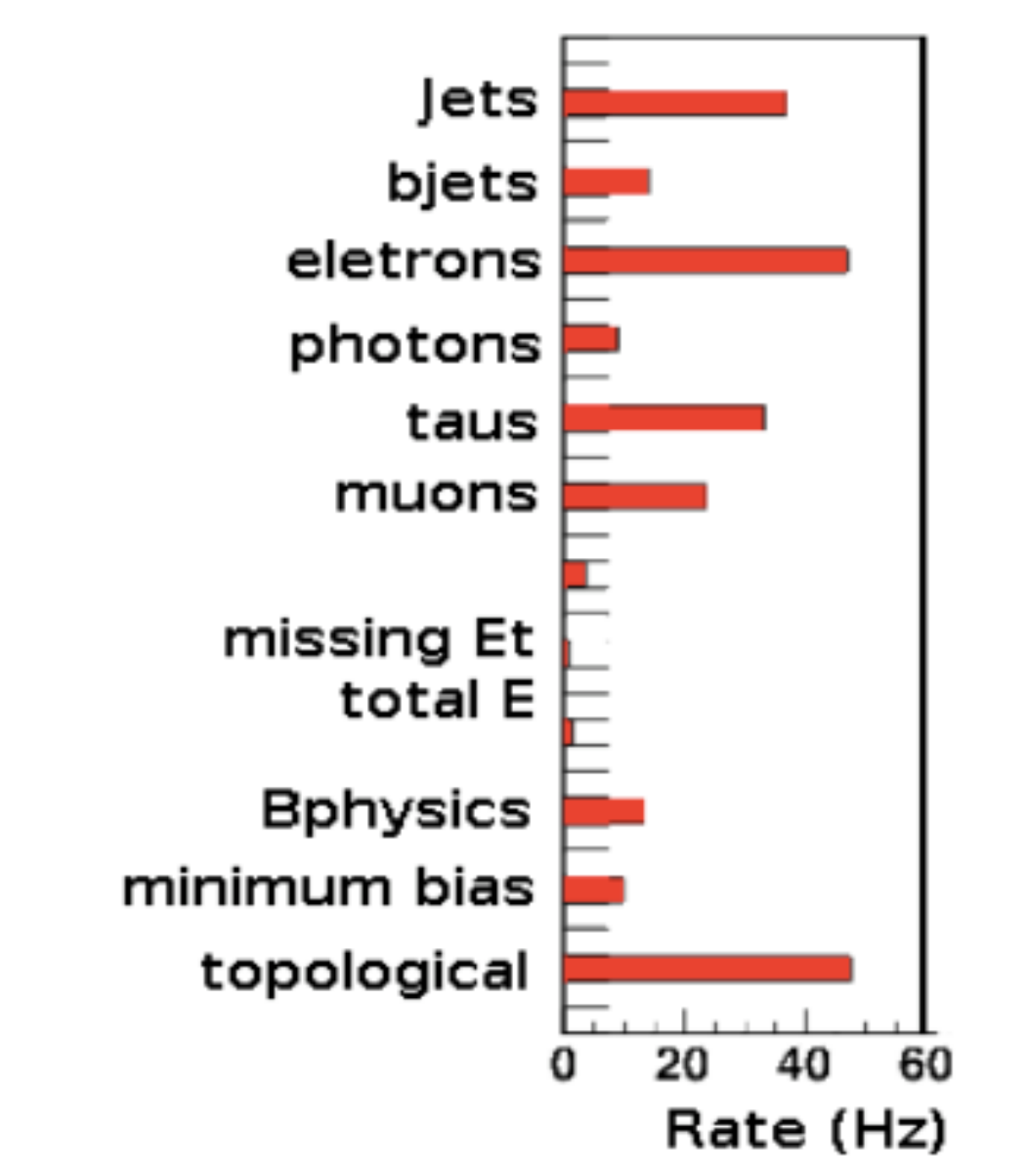}
\caption{Rate at the output of the EF for the example menu studied. 
The histogram shows inclusive rates. }
\label{fig:rate}
\end{figure}

The rates for a given menu is studied by running the trigger
selection on a sample of ``minimum bias'' events (these are events selected 
with the loosest possible trigger requirements and
which, therefore, represent
the main trigger background).  
About $70 mb^{-1}$ of such events were generated and
fed through the L1 trigger simulation and HLT processing. 
The composition of rates from some groups of chains is shown on the Fig.
\ref{fig:rate} \cite{AtlasDetPaper:2008}.

The rates for menus designed for a luminosity of $10^{31} cm^{-1}s^{-1}$ can be determined in this way. However, for higher luminosity menus 
the minimum bias events are enriched 
by samples with an understood bias such as di-jets in order to obtain sufficient statistics for the higher $p_t$ thresholds used in these menus.

\section{Plans for trigger commissioning}
So far, the trigger has been commissioned with simulated data. 
When the first LHC beam is available, this will
be used to ``time-in'' the detector (adjust for the signal propogation
delays within the various detector and trigger components). A
simple trigger menu will be used for this with more progresively 
more complex selections
being introduced later. The sequence of menus used for commissioning of the trigger will be  
as follows; initially
including only L1 in the selection, the HLT will be either excluded
 or included in a mode where is does not perform any selection.  
The L1 triggers will be
set in coincidence with the signal from near beam detectors.  Once a
coarse timing has been achieved the low $p_T$items from low luminosity
menu can be added to the menu and the HLT can be added in the mode
when the selection decision is evaluated but nevertheless all events 
are recorded.
With the increasing luminosity the timing can be fine-tuned and the
HLT selection turned on gradually  \cite{AtlasDetPaper:2008}.

When moving to higher luminosities the rate will need to be controlled by
tighter selections based on mid-$p_T$ items, disabling or highly prescalling
the low-$p_T$ thresholds and introducing new high-$p_T$ chains. As an 
alternative to removing or prescaling low -$p_T$ items, these items can be
required to have a higher multiplicity or new signatures can be formed from
the combination of one or more simple selection items. It should
be noted that more complex selections will be required at the  
nominal LHC luminosity where the rate of Standard-Model signatures will become 
impossible to record.

\section{Integration with the Data Acquisition System}
\begin{figure}[!ht]
\centering
\includegraphics[width=2.7in, height=2.1in]{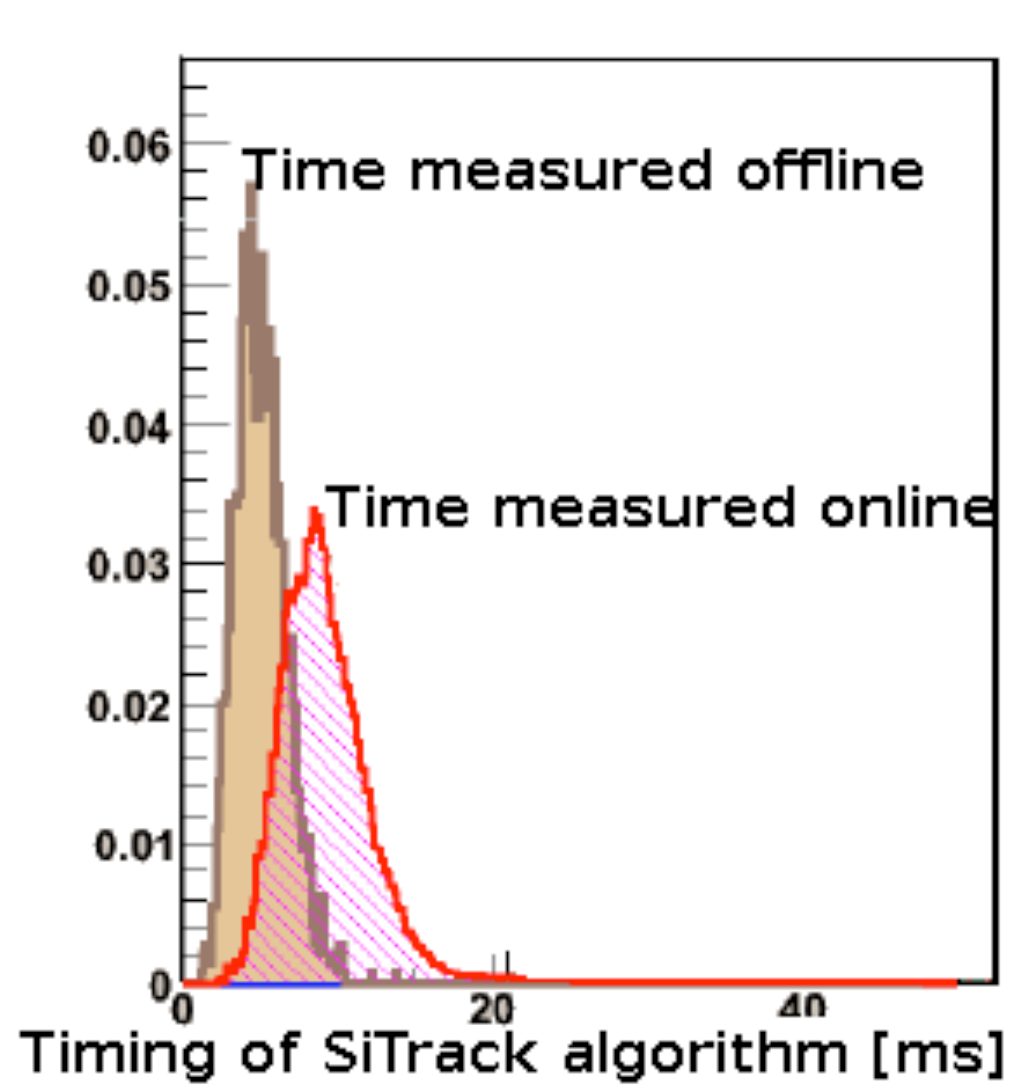}
\caption{The effect of data collection overhead seen during technical
  run compared to run on simulator. The ATLAS second level trigger is
  designed to request data on demand from detector readout
  buffers. This is so in order to postpone network intensive event
  building process until rejection is done by second trigger level.
  Request and delivery of the detector data adds to the time budget
  and therefore some triggering scenarios can be excluded due to this
  technical reason.}
\label{fig:online}
\end{figure}

In addition to the offline menu performance  studies,
 the HLT algorithms are taken to the
final L2 and EF farms and used with simulated data preloaded to the detector
readout system. In addition data taking periods are envisaged with detectors set to readout comics ray signals and with the trigger enabled. 
These tests help to understand the collective behaviors as well as
the long term trends of the system, such as those shown in Fig.~\ref{fig:online}.
Work is focussed arround two types of online tests; the so called
milestones runs using cosmic ray data from the real detectors and technical
runs where simulated data is downloaded to the front-end.  After each 
software release a series of
technical runs are performed and software fixes applied. The resulting 
certified releases are used for cosmic rays data taking. So far
a number of cycles of this kind have been performed.

\section{Addendum}
At the time of writing this report the trigger is involved in data taking with cosmic rays. The main purpose is detector commissioning.
The trigger is being used mainly to provide a streaming functionality. This means that the trigger is used to split the raw data into a number ofstreams for recording. The streaming decision is taken at the first level trigger and then preserved by both higher levels.
In addition to this, there is an active HLT selection performed based on tracking algorithms in order to enrich selected streams with the tracks useful for aligment of the ATLAS trackers. A part of the initial physics menu has also been deployed for testing purpose.

\bibliographystyle{plain}
\bibliography{ichep.bib}

\end{document}